\documentclass{ceurart}

\usepackage{acronym}
\usepackage{subcaption}
\usepackage{algorithm}
\usepackage{algpseudocode}
\usepackage{hyperref}
\usepackage{amsmath}
\usepackage{caption}
\usepackage{booktabs}
\usepackage{threeparttable}
\usepackage{bm}

\begin{document}

\copyrightyear{2024}
\copyrightclause{Copyright for this paper by its authors.\\
  Use permitted under Creative Commons License Attribution 4.0
  International (CC BY 4.0).}

\conference{CLEF 2024: Conference and Labs of the Evaluation Forum, September 9--12, 2024, Grenoble, France}

%\title{CRUISE on Using Quantum Computers for Feature Selection in Recommender Systems}
\title{CRUISE on Quantum Computing for Feature Selection in Recommender Systems}

\title[mode=sub]{Notebook for the QuantumCLEF Lab at CLEF 2024}

\author[]{Jiayang Niu}[%
email=s4068570@student.rmit.edu.au
]

\author[]{Jie Li}[%
email=hey.jieli@gmail.com
]

\author[]{Ke Deng}[%
email=ke.deng@rmit.edu.au
]

\author[]{Yongli Ren}[%
email=yongli.ren@rmit.edu.au
]

\address[]{School of Computing Technologies, RMIT University, Melbourne, Victoria 3000}

\begin{abstract}
Using Quantum Computers to solve problems in Recommender Systems that classical computers cannot address is a worthwhile research topic. In this paper, we use Quantum Annealers to address the feature selection problem in recommendation algorithms. This feature selection problem is a Quadratic Unconstrained Binary Optimization (QUBO) problem. By incorporating Counterfactual Analysis, we significantly improve the performance of the item-based KNN recommendation algorithm compared to using pure Mutual Information. Extensive experiments have demonstrated that the use of Counterfactual Analysis holds great promise for addressing such problems.
\end{abstract}


\maketitle

\section{Introduction}
\label{sec:introduction}

% Introduce the context, motivations, and goals of your project.

Collaborative filtering technology~\cite{su2009survey,lee2012comparative}, which predicts potential user-item interactions based on the patterns of user behavior and item characteristics, is widely applied in recommendation algorithms, Some well-known techniques in this field include matrix factorization methods~\cite{koenigstein2012efficient}, neighborhood-based methods~\cite{adeniyi2016automated}, deep learning approaches~\cite{hidasi2015session,vaswani2017attention}, graph-based techniques~\cite{wang2019neural,he2020lightgcn}, factorization machines~\cite{yuan2016lambdafm}, hybrid methods~\cite{adomavicius2005toward}, Bayesian methods~\cite{lopes2016efficient}, and large language models (LLMs)~\cite{yang2022gram}. However, collaborative filtering technology~\cite{su2009survey} heavily relies on the quality of data. For instance, using user profiles, item features, reviews, images, and other information can significantly improve the performance of recommendation algorithms, but in some cases, it can also decrease their performance. Therefore, it's critical to distinguish what information are useful for recommendations so as to help the the construction of efficient systems and reduction of energy consumption~\cite{marchesin2020focal,strubell2019energy, himeur2021survey, adomavicius2012impact}.  
Quantum computers, with its use of qubits and quantum effects like superposition, entanglement, and quantum tunneling, is an effective tool for identifying useful information from redundant data~\cite{lu2023quantum}. It significantly enhances the processing speed of search problems and large integer factorization~\cite{ferrari2022towards}. Therefore, in this paper, we aim to find useful features for recommendations by leveraging quantum computing techniques. Our goal is to improve the efficiency and accuracy of recommendation systems by identifying and utilizing relevant data, thereby reducing computational requirements and energy consumption~\cite{ferrari2022towards,glover2019quantum,pilato2022survey}.

In QuantumCLEF 2024, we focus on Task 1B, where 150 and 500 features are provided for each item, respectively\cite{clef-qclef:2024-ceur,clef-qclef:2024-lncs}. We will analyze these features to extract the most relevant ones for recommender systems. The task requires participants to use Quantum Annealing and Simulated Annealing to select appropriate features from the given data for an Item-Based KNN recommendation algorithm (\textbf{Item-KNN}). The organizers provided an example of feature selection by using Mutual Information~\cite{ferrari2022towards}. However, our preliminary experiments showed that using only Mutual Information for feature selection resulted in limited improvement in the performance of Item-KNN compared to using all features without any selection. This is because Mutual Information only reflects the mutual relationship between two variables and is not associated with the final goal of the recommendation algorithm. Therefore, to achieve better performance, we propose taking the impact of features on recommendation quality into consideration when performing feature selection. 
%explored other methods to enhance the feature selection capabilities of the Quantum Annealing and simulated annealing algorithms.

One approach to achieve this is through Counterfactual Analysis~\cite{pearl2016causal}, which is a causal research tool to examine the impact of a factor on the final result by hypothesizing the absence or alteration of that factor. This approach mainly considers three aspects: Which factors need to be evaluated? What metrics are used to assess the impact of these factors on the model's outcomes? And what models are used to derive the values of these metrics? In this work, due to the limited time for this task, we aim to measure and explore the impact of item features by Counterfactually Analyzing their effect on nDCG~\cite{jarvelin2002cumulated} performance of recommendation lists and we chose the KNN-based recommendation algorithm, a commonly used method in collaborative filtering, to perform these measurements. Specifically, we used Item-KNN to derive the change in nDCG values after removing a specific item feature. Since Mutual Information can reflect the relationship between two features, which may positively affects the final results, we did not discard it. Instead, we integrated the results of Counterfactual Analysis into Mutual Information using a temperature coefficient, which is used to control the influence of Counterfactual Analysis on the final results. Given the current limitations on the number of qubits in Quantum Computers, directly performing Quantum Annealing on 500 variables remains a challenging task. Therefore, in this task, we first partitioned the 500 features into subsets manageable by the Quantum Computer, and then combined the results.

The paper is organized as follows: Section~\ref{sec:related} introduces related works; Section~\ref{sec:methodology} describes the QUBO formulation, how Mutual Information is applied to QUBO for feature selection, and our proposed method of using Counterfactual Analysis for feature selection in QUBO; Section~\ref{sec:setup} explains our experimental setup and experimental result; Section~\ref{sec:results} discusses our main findings; finally, Section~\ref{sec:conclusion} draws some conclusions and outlooks for future work.

\section{Related Work}
\label{sec:related}
\subsection{Quantum Computers}
In recent years, the rapid development of Quantum Computers has demonstrated their tremendous potential in solving problems that Classical Computer cannot address, such as NP and NP-hard problems~\cite{bittel2021training}. Based on their functionality and application scenarios, Quantum Computers can be categorized into Universal Quantum Computers, Quantum Annealers, Quantum Machine Learning Accelerators, and others~\cite{gill2022quantum}. Recent studies have utilized Quantum Annealers for feature selection to enhance the performance of recommendation systems or retrieval systems~\cite{nembrini2021feature,nikitin2022quantum,ferrari2022towards}. Nembrini et al.~\cite{nembrini2021feature} attempted to apply Quantum Computers to recommendation systems by using Quantum Annealing to solve a hybrid feature selection approach. Their work demonstrates that current Quantum Computers are already capable of addressing real-world recommendation system problems. Nikitin et.al.\cite{nikitin2022quantum} reproduced Nembrini's work and employed Tensor Train-based Optimization (TTOpt) as an optimizer for the cold start problem in recommendation systems. MIQUBO~\cite{ferrari2022towards} discussed the problem of feature selection using Quantum Computers and formalizes it as a Quadratic Unconstrained Binary Optimization (QUBO) problem. It demonstrates the potential of Quantum Computers to solve ranking and classification problems more efficiently.

\subsection{Counterfactual Analysis}
Existing deep learning models have complex decision-making processes that are difficult for people to understand, often functioning as black-box models, Counterfactual Analysis is a highly effective method for helping people understand these complex models and robust them~\cite{verma2020counterfactual}. For example, $\textbf{CF}^2$~\cite{olson2021counterfactual} used Counterfactual Analysis to explore the explanations of Graph Neural Networks. In recommender systems, Counterfactual Analysis is primarily used for explainability and to combat data sparsity. ACCENT~\cite{tran2021counterfactual} was the first to apply Counterfactual Analysis to neural network-based recommendation algorithms. CountER~\cite{tan2021counterfactual} utilizes Counterfactual Analysis to construct a low-complexity, high-strength model for explaining recommendation systems. It also highlights that using Counterfactual Analysis contributes to the interpretability and evaluation of recommendation systems. Zhang et al~\cite{zhang2021causerec} designed a CauseRec framework that utilizes Counterfactual to enhance representations in the data distribution, aiming to mitigate data sparsity.

In summary, Counterfactual Analysis can help people understand complex deep learning decision systems and has the potential to analyze how various factors interact in recommendation systems. Given the current advancements in Quantum Computers, utilizing Counterfactual Analysis combined with the ability of Quantum Computers to handle NP problems presents a promising direction.

\section{Methodology}
\label{sec:methodology}
\subsection{Preliminary}
\subsubsection{QUBO Formulation}
In this work, we follow the approach described in~\cite{ferrari2022towards}, which utilizes Quantum Annealing for feature selection. To apply these methods, the feature selection problem is formulated as a Quadratic Unconstrained Binary Optimization (QUBO) problem. The QUBO formulation can be used to solve certain NP and NP-hard optimization problems and is defined as follows~\cite{ferrari2022towards}:
\begin{equation}
\min Y = x^T Q x, \label{eq:Equation1}
\end{equation}
where $x$ is a binary vector of length $m$, with each element of the vector being either 0 or 1. $Q$ is a symmetric matrix, where each element represents the relationship between the elements of $x$. $m$ denotes the number of features to be selected. In other words, the elements of vector $x$ indicate
whether the corresponding features are selected, and the elements in $Q$ influence the search direction of the function, determining feature selection.

\subsubsection{Feature Selection Based on Mutual Information}

Following~\cite{ferrari2022towards}, Mutual Information QUBO (MIQUBO) is a quadratic feature selection model based on Mutual Information. MIQUBO aims to maximize the Mutual Information, which measures the dependency between two variables, and the Conditional Mutual Information, which measures the dependency between two variables given a target variable, of the selected features. In this context, the matrix $Q$ in Equation \ref{eq:Equation1} is defined as:

\begin{equation}
Q_{ij} = \begin{cases}
-\text{CMI}(f_i; y \mid f_j) & \text{if } i \neq j \\
-\text{MI}(f_i; y) & \text{if } i = j,
\end{cases}
\end{equation}

where $\text{MI}(f_i; y)$ is the Mutual Information between feature $f_i$ and target feature $y$, and $\text{CMI}(f_i; y \mid f_j)$ is the Conditional Mutual Information between feature $f_i$ and target feature $y$ given feature $f_j$. Since QUBO formulation is used to find the minimum state, a negative sign is required before MI and CMI.

To control the number of selected features, a penalty term is added to Equation~\ref{eq:Equation1}, which is then transformed to: 
\begin{equation}
\min Y = x^T Q x + \left( \sum_{i=1}^{N} x_i - k \right)^2.
\end{equation}
This formula will be minimized when selecting $k$ features, this also following the descriptions in~\cite{ferrari2022towards}.

\subsection{Counterfactual Analysis}

To better identify features directly associated with recommendation performance, we integrate a widely used recommendation ranking metric into Mutual Information through Counterfactual Analysis.

\subsubsection{Counterfactual Analysis for Feature Selection}
Counterfactual Analysis~\cite{pearl2016causal} is usually used to examine the causal relationship between conditions, decisions, and outcomes by hypothesizing how the results of observed events would change if the conditions and decisions were altered. In the field of Recommender System, Counterfactual Analysis is often used for the interpretability of recommendation models, helping researchers enhance algorithm performance~\cite{tan2021counterfactual,zhang2021causerec}. Inspired by existing works~\cite{tan2021counterfactual,zhang2021causerec}, the impact of item features can be explored by excluding the corresponding feature and analyzing the difference in recommendation performance between the recommendation lists generated by the model with and without the corresponding feature.

In this work, we use the widely used Item-KNN recommendation algorithm, termed as model $G$, and employ the recommendation performance metric Normalized Discounted Cumulative Gain (nDCG)~\cite{jarvelin2002cumulated} for Counterfactual Analysis. nDCG is defined as:
\begin{equation}
\text{E}_i = \text{nDCG}_{G(\text{F})} - \text{nDCG}_{G(\text{F} \setminus {f_i})},
\end{equation}
where $E_i$ represents the change in the nDCG result of the recommendation model $G$ after removing the feature $f_i$. $\text{nDCG}_{G(\text{F})}$ represents the nDCG@10 value obtained by the $G$ using all item features set $F$, while $\text{nDCG}_{G(\text{F} \setminus {f_i})}$ represents the nDCG@10 value obtained by the $G$ using features set which is set $F$ removing feature $i$. It is important to note that $E_i$ ultimately reflects the impact of feature $i$ on the result. Since the final outcome is influenced by the interactions between all features, simply removing features with positive $E_i$ values does not yield the optimal feature selection solution.

When $E_i \geq 0$, it indicates that the algorithm's performance decreases after removing the feature $ i $. The extent of this decrease reflects the positive impact of this feature on the algorithm. Conversely, an increase in the value reflects the negative impact of this feature on the algorithm. We hypothesize that if the selected set of features is $set(F^*) $, the maximization the sum of $E_i$ ($i \in set(F^*)$), the maximization the performance improvement of the baseline algorithm. Since the QUBO problem is a minimization optimization problem, we redefine $Q$ as follows:
\begin{equation}
Q_{ij} = \begin{cases}
-CMI(f_i; y \mid f_j) & \text{if } i \neq j \\
-MI(f_i; y)-\lambda \text{E}_i  & \text{if } i = j
\end{cases}
\end{equation}
where $ \lambda $ is a coefficient used to control the influence of $E$ on the search results. The larger the value of $ \lambda $, the greater the influence of $E $ on the final results. The overall process of the above algorithm, which we refer to as Counterfactual Analysis QUBO (CAQUBO), is as follows in Algorithm~\ref{alg:pseudocode}.

\begin{algorithm}
\caption{Counterfactual Analysis QUBO}\label{alg:pseudocode}
\begin{algorithmic}[1]
\State Initialize variable set $\text{E}$, set $\text{F}$, $n\gets len(\text{F})$, $k$, $Q$, $\lambda$
\Procedure{Calculate $\text{E}_i$}{}
    \For{$f_i$ in $\text{F}$}
        \State $\text{F}^\text{'} \gets \text{F}$
        \State $\text{F}^\text{'}$.pop($f_i$)
        \State $\text{E}_i \gets $\text{G}(\text{F}) - $\text{G}(\text{F}^\text{'})$
    \EndFor
    \State \textbf{return} $\text{E}$
\EndProcedure
\Procedure{Feature Selection}{}
    \State Calculate MI and CMI
    \For{$f_i$ in $\text{F}$}
        \State $Q_{ii} = -\text{MI}(f_i; y) - \lambda \text{E}_i $
    \EndFor
    \For{$f_i$ in $\text{F}$}
        \For{$f_j$ in $\text{F}$}
            \State $Q_{ij} = -\text{CMI}(f_i; y \mid f_j)$
        \EndFor
    \EndFor
    \State set $\text{F}^* \gets$ QA or SA $\gets Q$ and $\lambda$ \# Input parameters $Q$ and $\lambda$ into the Quantum Annealer.
    \State \textbf{return} set $\text{F}^*$ \# Selected Feature Set
\EndProcedure
\end{algorithmic}
\end{algorithm}

\subsection{Handling Large Feature Set}
\label{sec:3.3}

Although Quantum Computers are developing rapidly, the limitation in the number of qubits restricts them to handling only a limited number of feature selection problems. For selecting from 500 features, we partition them into several subsets and use Quantum Annealing (QA) or Simulated Annealing (SA) to perform feature selection on these subsets individually, then combine the results.

First, partition the 500 features into $n$ subsets by order, ${S_1, S_2, \cdots, S_i, \cdots, S_n}$, where $S_i$ is the $i$-th subset of features, and $n$ is the number of subsets.
\begin{equation}
{S_1, S_2, \cdots, S_i, \cdots, S_n} = \text{divide(F)} 
\end{equation}
Then, use Quantum Annealing (QA) or Simulated Annealing (SA) to perform feature selection on each subset, and combine the results:
\begin{equation}
\tilde{S} = \bigcup_{i=1}^{n} \text{QA/SA}(S_i), 
\end{equation}
where $\tilde{S}$ is the final selected features set, represents each partitioned subset of features, and $\text{QA/SA}$ $(S\_i)$ represents the selected features from subset $ S_i $ using QA and SA. The final feature set is obtained by merging the selected features from all subsets.

\section{Experimental Setup}
\label{sec:setup}
\noindent\textbf{\textit{Datasets}}: In this work, two tasks are undertaken: the first involves selecting appropriate features from a set of 150 item features for training $G$, and the second involves selecting features from a set of 500 item features. Three data sets are provided for these tasks: 150\_ICM, 500\_ICM, and URM. The 150\_ICM and 500\_ICM contain item features, while the URM includes interaction data between 1,890 users and 18,022 interacted items.

\noindent\textbf{\textit{Experimental parameter setting}}: We used a self-implemented Item-KNN recommendation model based on the problem statement to calculate $E $. The interaction data was split into training and test sets in an 80:20 ratio. It is worth noting that calculating $ E $ is very time-consuming, so we only used a subset of items for the calculations. In the use of Quantum Annealing (QA) and Simulated Annealing(SA), the coefficient $\lambda$ significantly affects the features selected by QA and SA. Due to the limited usage time of the Quantum Annealer (QA), it is necessary to use Simulated Annealing (SA) to explore the effectiveness of the selected features under different parameters $\lambda$ and $k$ before using QA. In preliminary experiment, we attempt \textbf{[$\bm{\lambda}$: 0, 1e1, 1e3, 1e5, 1e7]}, \textbf{[$\textbf{k}$: 50, 100, 130, 140, 145] in Feature 150} and \textbf{[$\bm{\lambda}$: 0, 1e1, 1e3, 1e5, 1e7]}, \textbf{[$\textbf{k}$: 300, 350, 400, 450, 470] in Feature 500}. For the selection of 500 features, $\textbf{n}$ (is mentioned in Section \ref{sec:3.3}) is set to 5. The  preliminary experiment results can be found in Table \ref{tab:table1}.

\noindent\textbf{\textit{Repeated Calculations}}: Due to the heuristic nature of Simulated Annealing (SA) and Quantum Annealing (QA), the final results may vary even with fixed parameters. To mitigate this effect, we perform multiple iterations of QA and SA under the same parameters and select the final feature set via voting. For example, we repeated the experiment five times. $f_i$ was not included in $F^*$ in any of the five experiments, while $f_j$ was included in $F^*$ in four out of the five experiments. Therefore, the final submitted feature set $F^*$ does not include $f_i$ but includes $f_j$.

\begin{table}[h!]
    \caption{nDCG@10 for Feature 150 and Feature 500 datasets individually using SA-based feature selection, with different numbers of selected features $k$ and different coefficients $\lambda$.}. 
    \label{tab:table1}
    \centering
    \begin{tabular}{cccccccccccccc}
        \toprule
        k & 50 & 100 & 130 & 140 & 145 & & 300 & 350  & 400 & 450 & 470\\
        \cline{2-6}  \cline{8-12}
        $\lambda$ & \multicolumn{5}{c}{Feature 150 nDCG@10} & & \multicolumn{5}{c}{Feature 500 nDCG@10} \\
        \midrule
         0   & 0.0602 & 0.0870 & 0.0968 & 0.1033 & 0.1018 & &  0.1078 &  0.0894 &  0.0971 &  0.0969 &  0.0991\\
         1   & 0.0870 & 0.0974 & 0.0999 & 0.1009 & 0.1029 & &  0.1066&  0.1108&  0.1195&  0.1291&  0.1197\\
         1e3 & 0.0755 & 0.1051 & 0.1151 & 0.1119 & 0.1152 & &  0.1206&  0.1249&  0.1257&  0.1305&  0.1302\\
         1e5 & \textbf{0.0878} & \textbf{0.1160} & \textbf{0.1232} & 0.1256 & \textbf{0.1180} & &  0.1224&  \textbf{0.1238}&  \textbf{0.1303}&  0.1290&  \textbf{0.1307}\\
         1e7 & 0.0795 & 0.1155 & 0.1221 & \textbf{0.1264} & \textbf{0.1180} & &  \textbf{0.1235}&  0.1218&  0.1298&  \textbf{0.1306}&  0.1293 \\
         \cline{2-6}  \cline{8-12}
         & \multicolumn{5}{l}{\textbf{150 Feature nDCG 0.1028}} & & \multicolumn{5}{l}{\textbf{500 Feature nDCG 0.0988}} \\
        \bottomrule
    \end{tabular}
\end{table}

\begin{table}[h!]
    \caption{This table contains the final data submitted to the organizers, with data sourced from the organizers' $\text{website}^\textbf{1}$. Due to the fact that when $\bm{\lambda}$ is too large, the values of elements in $\textbf{Q}$ become excessively large, which is detrimental to the performance of QA and SA, a coefficient $\bm{\mu}$ is applied to all elements in $\textbf{Q}$. An asterisk (\textbf{*}) after the sub\_ID indicates that the selected features are the result of repeated calculations. Those submissions was repeated five times to determine the final feature set.}
    \label{tab:table2}
    \begin{threeparttable}
    \centering
    \begin{tabular}{llllll}
    \toprule
    \textbf{150 Feature submissions} & \multicolumn{3}{l}{\textbf{All Feature nDCG 0.0810}}    &  &  \\
    \midrule
    \textbf{Parameters set} & \textbf{nDCG@10} & \textbf{Annealing Time} & \textbf{Type} & \textbf{nº features} & \textbf{sub\_id} \\
    \midrule
     $\text{k}$=140 ${\lambda}$=1e7 $\mu$=1e-5 & 0.0805 & 536250 & Q & 138 & 1\\
     $\text{k}$=140 ${\lambda}$=1e7 $\mu$=1e-3 & 0.0826 & 528844 & Q & 136 & 2\\
     $\text{k}$=140 ${\lambda}$=1e7 $\mu$=1e-3 & 0.0690 & 530804 & Q & 132 & 3\\
     $\text{k}$=140 ${\lambda}$=0 $\mu$=1  & 0.0763 & 558321 & Q & 133 & 4\\
     $\text{k}$=140 ${\lambda}$=1e7 $\mu$=1e-2 & 0.1003 & 1375068 & Q & 144 & $\text{5}^{*}$\\
     $\text{k}$=140 ${\lambda}$=1e7 $\mu$=1e-5 & 0.0998 & 1745487 & S & 140 & 1\\
     $\text{k}$=140 ${\lambda}$=1e7 $\mu$=1e-3 & 0.0993 & 17357899 & S & 140 & 2\\
     $\text{k}$=140 ${\lambda}$=1e7 $\mu$=1e-3 & 0.1001 & 1760252 & S & 140 & 3\\
     $\text{k}$=140 ${\lambda}$=0 $\mu$=1  & 0.0793 & 17387227 & S & 140 & 4\\
     $\text{k}$=140 ${\lambda}$=1e7 $\mu$=1e-2 & 0.1003 & 88395437 & S & 144 & $\text{5}^{*}$\\
    \toprule
    \textbf{500 Feature submissions} & \multicolumn{3}{l}{\textbf{All Feature nDCG 0.0827}}   &  &  \\
    \midrule
    $\text{k}$=450 ${\lambda}$=1e7 $\mu$=1e-2  & 0.0757 & 2287019 & Q & 407 & 1 \\ 
    $\text{k}$=450 ${\lambda}$=1e1 $\mu$=1  & 0.0839 & 2122701 & Q & 397 & 2 \\ 
    $\text{k}$=450 ${\lambda}$=1e7 $\mu$=1e-2   & 0.1196 & 43339285 & S & 450 & 1 \\ 
    $\text{k}$=450 ${\lambda}$=1e1 $\mu$=1  & 0.1198 & 42776695 & S & 450 & 2 \\ 
    \midrule
    \end{tabular}
    \begin{tablenotes}
    \footnotesize
    \item[\textbf{1}] \textbf{https://qclef.dei.unipd.it/clef2024-results.html}
    \end{tablenotes}
    \end{threeparttable}
\end{table}

\section{Results}
\label{sec:results}

Table \ref{tab:table1} describes the performance in nDCG@10 of $G$ using features selected by QA and SA under different parameters $\lambda$ and $k$. When $\lambda = 0$, QA and SA select features based solely on Mutual Information (MI) and Conditional Mutual Information (CMI). Across different values of parameter $k$, the performance of selected features in  $G$ rarely surpasses the performance in Counterfactual Analysis QUBO. As the parameter $\lambda$ increases, the performance of the features selected by QA and SA in the item-KNN shows significant improvement compared to using all features. The effectiveness of feature selection shows no significant improvement when $\lambda> 1e5 $ . This may be because as the value of $\lambda$ increases, the impact of MI and CMI on feature selection diminishes, causing QA and SA to rely entirely on $E$ for feature selection.

Table \ref{tab:table2} reflects the same situation: feature selection relying solely on MI and CMI does not surpass the performance in Counterfactual Analysis QUBO. After incorporating the counterfactual analysis-derived $E$ into \(Q\), the features selected by QA and SA show a significant performance improvement in item-KNN compared to using all features. An unusual observation is that, under the same parameters, the features selected by QA generally do not perform as well as those selected by SA in item-KNN, and sometimes do not even surpass the performance of using all features. During the experiments, we noticed that this is due to QA often returning results before finding the optimal solution.

\section{Conclusions and Future Work}
\label{sec:conclusion}

In this paper, we present the explorations conducted by our team and the details of our final submission for the QuantumCLEF 2024 activities. We used Counterfactual Analysis of individual item features to select appropriate features for item-KNN using Quantum Annealing. Our preliminary experiments and the results returned by QuantumCLEF 2024 demonstrated that our use of Counterfactual Analysis significantly improved the performance of item-KNN.

Within the limited time of QuantumCLEF, we attempted Counterfactual Analysis of individual features. However, because the performance of collaborative filtering is actually the result of feature interactions, Counterfactual Analysis of individual features has significant limitations. Additionally, since Quantum Annealing cannot directly handle the selection of 500 features, we adopted a sequential partitioning and merging approach. As negative features are not uniformly distributed by their indices among all features, this sequential partitioning and merging method still requires improvement.

\bibliography{bibliography}

\acrodef{3G}[3G]{Third Generation Mobile System}
\acrodef{5S}[5S]{Streams, Structures, Spaces, Scenarios, Societies}
\acrodef{AA}[AA]{Active Agreements}
\acrodef{AAAI}[AAAI]{Association for the Advancement of Artificial Intelligence}
\acrodef{AAL}[AAL]{Annotation Abstraction Layer}
\acrodef{AAM}[AAM]{Automatic Annotation Manager}
\acrodef{AAP}[AAP]{Average Average Precision}
\acrodef{ACLIA}[ACLIA]{Advanced Cross-Lingual Information Access}
\acrodef{ACM}[ACM]{Association for Computing Machinery}
\acrodef{AD}[AD]{Active Disagreements}
\acrodef{ADSL}[ADSL]{Asymmetric Digital Subscriber Line}
\acrodef{ADUI}[ADUI]{ADministrator User Interface}
\acrodef{AIP}[AIP]{Archival Information Package}
\acrodef{AJAX}[AJAX]{Asynchronous JavaScript Technology and \acs{XML}}
\acrodef{ALU}[ALU]{Aritmetic-Logic Unit}
\acrodef{AMUSID}[AMUSID]{Adaptive MUSeological IDentity-service}
\acrodef{ANOVA}[ANOVA]{ANalysis Of VAriance}
\acrodef{ANSI}[ANSI]{American National Standards Institute}
\acrodef{AP}[AP]{Average Precision}
\acrodef{APC}[APC]{AP Correlation}
\acrodef{API}[API]{Application Program Interface}
\acrodef{AR}[AR]{Address Register}
\acrodef{AS}[AS]{Annotation Service}
\acrodef{ASAP}[ASAP]{Adaptable Software Architecture Performance}
\acrodef{ASI}[ASI]{Annotation Service Integrator}
\acrodef{ASL}[ASL]{Achieved Significance Level}
\acrodef{ASM}[ASM]{Annotation Storing Manager}
\acrodef{ASR}[ASR]{Automatic Speech Recognition}
\acrodef{ASUI}[ASUI]{ASsessor User Interface}
\acrodef{ATIM}[ATIM]{Annotation Textual Indexing Manager}
\acrodef{AUC}[AUC]{Area Under the ROC Curve}
\acrodef{AUI}[AUI]{Administrative User Interface}
\acrodef{AWARE}[AWARE]{Assessor-driven Weighted Averages for Retrieval Evaluation}
\acrodef{BANKS-I}[BANKS-I]{Browsing ANd Keyword Searching I}
\acrodef{BANKS-II}[BANKS-II]{Browsing ANd Keyword Searching II}
\acrodef{BH}[BH]{Benjamini-Hochberg}
\acrodef{bpref}[bpref]{Binary Preference}
\acrodef{BNF}[BNF]{Backus and Naur Form}
\acrodef{BPM}[BPM]{Bejeweled Player Model}
\acrodef{BRICKS}[BRICKS]{Building Resources for Integrated Cultural Knowledge Services}
\acrodef{CAN}[CAN]{Content Addressable Netword}
\acrodef{CAS}[CAS]{Content-And-Structure}
\acrodef{CBSD}[CBSD]{Component-Based Software Developlement}
\acrodef{CBSE}[CBSE]{Component-Based Software Engineering}
\acrodef{CB-SPE}[CB-SPE]{Component-Based \acs{SPE}}
\acrodef{CD}[CD]{Collaboration Diagram}
\acrodef{CD}[CD]{Compact Disk}
\acrodef{CDF}[CDF]{Cumulative Density Function}
\acrodef{CENL}[CENL]{Conference of European National Librarians}
\acrodef{CIDOC CRM}[CIDOC CRM]{CIDOC Conceptual Reference Model}
\acrodef{CIR}[CIR]{Current Instruction Register}
\acrodef{CIRCO}[CIRCO]{Coordinated Information Retrieval Components Orchestration}
\acrodef{CG}[CG]{Cumulated Gain}
\acrodef{CL}[CL]{Curriculum Learning}
\acrodef{CL-ESA}[CL-ESA]{Cross-Lingual Explicit Semantic Analysis}
\acrodef{CLAIRE}[CLAIRE]{Combinatorial visuaL Analytics system for Information Retrieval Evaluation}
\acrodef{CLEF1}[CLEF]{Cross-Language Evaluation Forum}
\acrodef{CLEF}[CLEF]{Conference and Labs of the Evaluation Forum}
\acrodef{CLIR}[CLIR]{Cross Language Information Retrieval}
\acrodef{CM}[CM]{Continuation Methods}
\acrodef{CMS}[CMS]{Content Management System}
\acrodef{CMT}[CMT]{Campaign Management Tool}
\acrodef{CNR}[CNR]{Italian National Council of Research}
\acrodef{CO}[CO]{Content-Only}
\acrodef{COD}[COD]{Code On Demand}
\acrodef{CODATA}[CODATA]{Committee on Data for Science and Technology}
\acrodef{COLLATE}[COLLATE]{Collaboratory for Annotation Indexing and Retrieval of Digitized Historical Archive Material}
\acrodef{CP}[CP]{Characteristic Pattern}
\acrodef{CPE}[CPE]{Control Processor Element}
\acrodef{CPU}[CPU]{Central Processing Unit}
\acrodef{CQL}[CQL]{Contextual Query Language}
\acrodef{CRP}[CRP]{Cumulated Relative Position}
\acrodef{CRUD}[CRUD]{Create--Read--Update--Delete}
\acrodef{CS}[CS]{Characteristic Structure}
\acrodef{CSM}[CSM]{Campaign Storing Manager}
\acrodef{CSS}[CSS]{Cascading Style Sheets}
\acrodef{CTR}[CTR]{Click-Through Rate}
\acrodef{CU}[CU]{Control Unit}
\acrodef{CUI}[CUI]{Client User Interface}
\acrodef{CV}[CV]{Cross-Validation}
\acrodef{DAFFODIL}[DAFFODIL]{Distributed Agents for User-Friendly Access of Digital Libraries}
\acrodef{DAO}[DAO]{Data Access Object}
\acrodef{DARE}[DARE]{Drawing Adequate REpresentations}
\acrodef{DARPA}[DARPA]{Defense Advanced Research Projects Agency}
\acrodef{DAS}[DAS]{Distributed Annotation System}
\acrodef{DB}[DB]{DataBase}
\acrodef{DBMS}[DBMS]{DataBase Management System}
\acrodef{DC}[DC]{Dublin Core}
\acrodef{DCG}[DCG]{Discounted Cumulated Gain}
\acrodef{DCMI}[DCMI]{Dublin Core Metadata Initiative}
\acrodef{DCV}[DCV]{Document Cut--off Value}
\acrodef{DD}[DD]{Deployment Diagram}
\acrodef{DDC}[DDC]{Dewey Decimal Classification}
\acrodef{DDS}[DDS]{Direct Data Structure}
\acrodef{DF}[DF]{Degrees of Freedom}
\acrodef{DFI}[DFI]{Divergence From Independence}
\acrodef{DFR}[DFR]{Divergence From Randomness}
\acrodef{DHT}[DHT]{Distributed Hash Table}
\acrodef{DI}[DI]{Digital Image}
\acrodef{DIKW}[DIKW]{Data, Information, Knowledge, Wisdom}
\acrodef{DIL}[DIL]{\acs{DIRECT} Integration Layer}
\acrodef{DiLAS}[DiLAS]{Digital Library Annotation Service}
\acrodef{DIRECT}[DIRECT]{Distributed Information Retrieval Evaluation Campaign Tool}
\acrodef{DKMS}[DKMS]{Data and Knowledge Management System}
\acrodef{DL}[DL]{Digital Library}
\acrodefplural{DL}[DL]{Digital Libraries}
\acrodef{DLMS}[DLMS]{Digital Library Management System}
\acrodef{DLOG}[DL]{Description Logics}
\acrodef{DLS}[DLS]{Digital Library System}
\acrodef{DLSS}[DLSS]{Digital Library Service System}
\acrodef{DM}[DM]{Data Mining}
\acrodef{DO}[DO]{Digital Object}
\acrodef{DOI}[DOI]{Digital Object Identifier}
\acrodef{DOM}[DOM]{Document Object Model}
\acrodef{DoMDL}[DoMDL]{Document Model for Digital Libraries}
\acrodef{DP}[DP]{Discriminative Power}
\acrodef{DPBF}[DPBF]{Dynamic Programming Best-First}
\acrodef{DR}[DR]{Data Register}
\acrodef{DRIVER}[DRIVER]{Digital Repository Infrastructure Vision for European Research}
\acrodef{DTD}[DTD]{Document Type Definition}
\acrodef{DVD}[DVD]{Digital Versatile Disk}
\acrodef{EAC-CPF}[EAC-CPF]{Encoded Archival Context for Corporate Bodies, Persons, and Families}
\acrodef{EAD}[EAD]{Encoded Archival Description}
\acrodef{EAN}[EAN]{International Article Number}
\acrodef{EBU}[EBU]{Expected Browsing Utility}
\acrodef{ECD}[ECD]{Enhanced Contenty Delivery}
\acrodef{ECDL}[ECDL]{European Conference on Research and Advanced Technology for Digital Libraries}
\acrodef{EDM}[EDM]{Europeana Data Model}
\acrodef{EG}[EG]{Execution Graph}
\acrodef{ELDA}[ELDA]{Evaluation and Language resources Distribution Agency}
\acrodef{ELRA}[ELRA]{European Language Resources Association}
\acrodef{EM}[EM]{Expectation Maximization}
\acrodef{EMMA}[EMMA]{Extensible MultiModal Annotation}
\acrodef{EPROM}[EPROM]{Erasable Programmable \acs{ROM}}
\acrodef{EQNM}[EQNM]{Extended Queueing Network Model}
\acrodef{ER}[ER]{Entity--Relationship}
\acrodef{ERR}[ERR]{Expected Reciprocal Rank}
\acrodef{ERS}[ERS]{Empirical Relational System}
\acrodef{ESA}[ESA]{Explicit Semantic Analysis}
\acrodef{ESL}[ESL]{Expected Search Length}
\acrodef{ETL}[ETL]{Extract-Transform-Load}
\acrodef{FAST}[FAST]{Flexible Annotation Service Tool}
\acrodef{FDR}[FDR]{False Discovery Rate}
\acrodef{FIFO}[FIFO]{First-In / First-Out}
\acrodef{FIRE}[FIRE]{Forum for Information Retrieval Evaluation}
\acrodef{FN}[FN]{False Negative}
\acrodef{FNR}[FNR]{False Negative Rate}
\acrodef{FOAF}[FOAF]{Friend of a Friend}
\acrodef{FORESEE}[FORESEE]{FOod REcommentation sErvER}
\acrodef{FP}[FP]{False Positive}
\acrodef{FPR}[FPR]{False Positive Rate}
\acrodef{FWER}[FWER]{Family-wise Error Rate}
\acrodef{GIF}[GIF]{Graphics Interchange Format}
\acrodef{GIR}[GIR]{Geografic Information Retrieval}
\acrodef{GAP}[GAP]{Graded Average Precision}
\acrodef{GLM}[GLM]{General Linear Model}
\acrodef{GLMM}[GLMM]{General Linear Mixed Model}
\acrodef{GMAP}[GMAP]{Geometric Mean Average Precision}
\acrodef{GoP}[GoP]{Grid of Points}
\acrodef{GPRS}[GPRS]{General Packet Radio Service}
\acrodef{gP}[gP]{Generalized Precision}
\acrodef{gR}[gR]{Generalized Recall}
\acrodef{gRBP}[gRBP]{Graded Rank-Biased Precision}
\acrodef{GT}[GT]{Generalizability Theory}
\acrodef{GTIN}[GTIN]{Global Trade Item Number}
\acrodef{GUI}[GUI]{Graphical User Interface}
\acrodef{GW}[GW]{Gateway}
\acrodef{HCI}[HCI]{Human Computer Interaction}
\acrodef{HDS}[HDS]{Hybrid Data Structure}
\acrodef{HIR}[HIR]{Hypertext Information Retrieval}
\acrodef{HIT}[HIT]{Human Intelligent Task}
\acrodef{HITS}[HITS]{Hyperlink-Induced Topic Search}
\acrodef{HMM}[HMM]{Hidden Markov Model}
\acrodef{HTML}[HTML]{HyperText Markup Language}
\acrodef{HTTP}[HTTP]{HyperText Transfer Protocol}
\acrodef{HSD}[HSD]{Honestly Significant Difference}
\acrodef{ICA}[ICA]{International Council on Archives}
\acrodef{ICSU}[ICSU]{International Council for Science}
\acrodef{IDF}[IDF]{Inverse Document Frequency}
\acrodef{IDS}[IDS]{Inverse Data Structure}
\acrodef{IEEE}[IEEE]{Institute of Electrical and Electronics Engineers}
\acrodef{IEI}[IEI]{Istituto della Enciclopedia Italiana fondata da Giovanni Treccani}
\acrodef{IETF}[IETF]{Internet Engineering Task Force}
\acrodef{IIR}[IIR]{Interactive Information Retrieval}
\acrodef{IMS}[IMS]{Information Management System}
\acrodef{IMSPD}[IMS]{Information Management Systems Research Group}
\acrodef{indAP}[indAP]{Induced Average Precision}
\acrodef{infAP}[infAP]{Inferred Average Precision}
\acrodef{INEX}[INEX]{INitiative for the Evaluation of \acs{XML} Retrieval}
\acrodef{INS-M}[INS-M]{Inverse Set Data Model}
\acrodef{INTR}[INTR]{Interrupt Register}
\acrodef{IP}[IP]{Internet Protocol}
\acrodef{IPSA}[IPSA]{Imaginum Patavinae Scientiae Archivum}
\acrodef{IR}[IR]{Information Retrieval}
\acrodef{IRON}[IRON]{Information Retrieval ON}
\acrodef{IRON2}[IRON$^2$]{Information Retrieval On aNNotations}
\acrodef{IRON-SAT}[IRON-SAT]{\acs{IRON} - Statistical Analysis Tool}
\acrodef{IRS}[IRS]{Information Retrieval System}
\acrodef{ISAD(G)}[ISAD(G)]{International Standard for Archival Description (General)}
\acrodef{ISBN}[ISBN]{International Standard Book Number}
\acrodef{ISIS}[ISIS]{Interactive SImilarity Search}
\acrodef{ISJ}[ISJ]{Interactive Searching and Judging}
\acrodef{ISO}[ISO]{International Organization for Standardization}
\acrodef{ITU}[ITU]{International Telecommunication Union }
\acrodef{ITU-T}[ITU-T]{Telecommunication Standardization Sector of \acs{ITU}}
\acrodef{IV}[IV]{Information Visualization}
\acrodef{JAN}[JAN]{Japanese Article Number}
\acrodef{JDBC}[JDBC]{Java DataBase Connectivity}
\acrodef{JMB}[JMB]{Java--Matlab Bridge}
\acrodef{JPEG}[JPEG]{Joint Photographic Experts Group}
\acrodef{JSON}[JSON]{JavaScript Object Notation}
\acrodef{JSP}[JSP]{Java Server Pages}
\acrodef{JTE}[JTE]{Java-Treceval Engine}
\acrodef{KDE}[KDE]{Kernel Density Estimation}
\acrodef{KLD}[KLD]{Kullback-Leibler Divergence}
\acrodef{KLAPER}[KLAPER]{Kernel LAnguage for PErformance and Reliability analysis}
\acrodef{LAM}[LAM]{Libraries, Archives, and Museums}
\acrodef{LAM2}[LAM]{Logistic Average Misclassification}
\acrodef{LAN}[LAN]{Local Area Network}
\acrodef{LD}[LD]{Linked Data}
\acrodef{LEAF}[LEAF]{Linking and Exploring Authority Files}
\acrodef{LIDO}[LIDO]{Lightweight Information Describing Objects}
\acrodef{LIFO}[LIFO]{Last-In / First-Out}
\acrodef{LM}[LM]{Language Model}
\acrodef{LMT}[LMT]{Log Management Tool}
\acrodef{LOD}[LOD]{Linked Open Data}
\acrodef{LODE}[LODE]{Linking Open Descriptions of Events}
\acrodef{LpO}[LpO]{Leave-$p$-Out}
\acrodef{LRM}[LRM]{Local Relational Model}
\acrodef{LRU}[LRU]{Last Recently Used}
\acrodef{LS}[LS]{Lexical Signature}
\acrodef{LSM}[LSM]{Log Storing Manager}
\acrodef{LtR}[LtR]{Learning to Rank}
\acrodef{LUG}[LUG]{Lexical Unit Generator}
\acrodef{MA}[MA]{Mobile Agent}
\acrodef{MA}[MA]{Moving Average}
\acrodef{MACS}[MACS]{Multilingual ACcess to Subjects}
\acrodef{MADCOW}[MADCOW]{Multimedia Annotation of Digital Content Over the Web}
\acrodef{MAD}[MAD]{Mean Assessed Documents}
\acrodef{MADP}[MADP]{Mean Assessed Documents Precision}
\acrodef{MADS}[MADS]{Metadata Authority Description Standard}
\acrodef{MAP}[MAP]{Mean Average Precision}
\acrodef{MARC}[MARC]{Machine Readable Cataloging}
\acrodef{MATTERS}[MATTERS]{MATlab Toolkit for Evaluation of information Retrieval Systems}
\acrodef{MDA}[MDA]{Model Driven Architecture}
\acrodef{MDD}[MDD]{Model-Driven Development}
\acrodef{METS}[METS]{Metadata Encoding and Transmission Standard}
\acrodef{MIDI}[MIDI]{Musical Instrument Digital Interface}
\acrodef{MIME}[MIME]{Multipurpose Internet Mail Extensions}
\acrodef{ML}[ML]{Machine Learning}
\acrodef{MLE}[MLE]{Maximum Likelihood Estimation}
\acrodef{MLIA}[MLIA]{MultiLingual Information Access}
\acrodef{MM}[MM]{Machinery Model}
\acrodef{MMU}[MMU]{Memory Management Unit}
\acrodef{MODS}[MODS]{Metadata Object Description Schema}
\acrodef{MOF}[MOF]{Meta-Object Facility}
\acrodef{MP}[MP]{Markov Precision}
\acrodef{MPEG}[MPEG]{Motion Picture Experts Group}
\acrodef{MRD}[MRD]{Machine Readable Dictionary}
\acrodef{MRF}[MRF]{Markov Random Field}
\acrodef{MRR}[MRR]{Mean Reciprocal Rank}
\acrodef{MS}[MS]{Mean Squares}
\acrodef{MSAC}[MSAC]{Multilingual Subject Access to Catalogues}
\acrodef{MSE}[MSE]{Mean Square Error}
\acrodef{MT}[MT]{Machine Translation}
\acrodef{MV}[MV]{Majority Vote}
\acrodef{MVC}[MVC]{Model-View-Controller}
\acrodef{NACSIS}[NACSIS]{NAtional Center for Science Information Systems}
\acrodef{NAP}[NAP]{Network processors Applications Profile}
\acrodef{NCP}[NCP]{Normalized Cumulative Precision}
\acrodef{nCG}[nCG]{Normalized Cumulated Gain}
\acrodef{nCRP}[nCRP]{Normalized Cumulated Relative Position}
\acrodef{nDCG}[nDCG]{Normalized Discounted Cumulated Gain}
\acrodef{nMCG}[nMCG]{Normalized Markov Cumulated Gain}
\acrodef{NESTOR}[NESTOR]{NEsted SeTs for Object hieRarchies}
\acrodef{NEXI}[NEXI]{Narrowed Extended XPath I}
\acrodef{NII}[NII]{National Institute of Informatics}
\acrodef{NISO}[NISO]{National Information Standards Organization}
\acrodef{NIST}[NIST]{National Institute of Standards and Technology}
\acrodef{NLP}[NLP]{Natural Language Processing}
\acrodef{NN}[NN]{Neural Network}
\acrodef{NP}[NP]{Network Processor}
\acrodef{NR}[NR]{Normalized Recall}
\acrodef{NRS}[NRS]{Numerical Relational System}
\acrodef{NS-M}[NS-M]{Nested Set Model}
\acrodef{NTCIR}[NTCIR]{NII Testbeds and Community for Information access Research}
\acrodef{OAI}[OAI]{Open Archives Initiative}
\acrodef{OAI-ORE}[OAI-ORE]{Open Archives Initiative Object Reuse and Exchange}
\acrodef{OAI-PMH}[OAI-PMH]{Open Archives Initiative Protocol for Metadata Harvesting}
\acrodef{OAIS}[OAIS]{Open Archival Information System}
\acrodef{OC}[OC]{Operation Code}
\acrodef{OCLC}[OCLC]{Online Computer Library Center}
\acrodef{OMG}[OMG]{Object Management Group}
\acrodef{OO}[OO]{Object Oriented}
\acrodef{OODB}[OODB]{Object-Oriented \acs{DB}}
\acrodef{OODBMS}[OODBMS]{Object-Oriented \acs{DBMS}}
\acrodef{OPAC}[OPAC]{Online Public Access Catalog}
\acrodef{OQL}[OQL]{Object Query Language}
\acrodef{ORP}[ORP]{Open Relevance Project}
\acrodef{OSIRIS}[OSIRIS]{Open Service Infrastructure for Reliable and Integrated process Support}
\acrodef{P}[P]{Precision}
\acrodef{P2P}[P2P]{Peer-To-Peer}
\acrodef{PA}[PA]{Passive Agreements}
\acrodef{PAMT}[PAMT]{Pool-Assessment Management Tool}
\acrodef{PASM}[PASM]{Pool-Assessment Storing Manager}
\acrodef{PC}[PC]{Program Counter}
\acrodef{PCP}[PCP]{Pre-Commercial Procurement}
\acrodef{PCR}[PCR]{Peripherical Command Register}
\acrodef{PD}[PD]{Passive Disagreements}
\acrodef{PDA}[PDA]{Personal Digital Assistant}
\acrodef{PDF}[PDF]{Probability Density Function}
\acrodef{PDR}[PDR]{Peripherical Data Register}
\acrodef{PIR}[PIR]{Personalized Information Retrieval}
\acrodef{POI}[POI]{\acs{PURL}-based Object Identifier}
\acrodef{PoS}[PoS]{Part of Speech}
\acrodef{PAA}[PAA]{Proportion of Active Agreements}
\acrodef{PPA}[PPA]{Proportion of Passive Agreements}
\acrodef{PPE}[PPE]{Programmable Processing Engine}
\acrodef{PREFORMA}[PREFORMA]{PREservation FORMAts for culture information/e-archives}
\acrodef{PRIMAD}[PRIMAD]{Platform, Research goal, Implementation, Method, Actor, and Data}
\acrodef{PRIMAmob-UML}[PRIMAmob-UML]{mobile \acs{PRIMA-UML}}
\acrodef{PRIMA-UML}[PRIMA-UML]{PeRformance IncreMental vAlidation in \acs{UML}}
\acrodef{PROM}[PROM]{Programmable \acs{ROM}}
\acrodef{PROMISE}[PROMISE]{Participative Research labOratory  for Multimedia and Multilingual Information Systems Evaluation}
\acrodef{pSQL}[pSQL]{propagate \acs{SQL}}
\acrodef{PUI}[PUI]{Participant User Interface}
\acrodef{PURL}[PURL]{Persistent \acs{URL}}
\acrodef{QA}[QA]{Question Answering}
\acrodef{QE}[QE]{Query Expansion}
\acrodef{QoS-UML}[QoS-UML]{\acs{UML} Profile for QoS and Fault Tolerance}
\acrodef{QPA}[QPA]{Query Performance Analyzer}
\acrodef{QPP}[QPP]{Query Performance Prediction}
\acrodef{R}[R]{Recall}
\acrodef{RAM}[RAM]{Random Access Memory}
\acrodef{RAMM}[RAM]{Random Access Machine}
\acrodef{RBO}[RBO]{Rank-Biased Overlap}
\acrodef{RBP}[RBP]{Rank-Biased Precision}
\acrodef{RBTO}[RBTO]{Rank-Based Total Order}
\acrodef{RDBMS}[RDBMS]{Relational \acs{DBMS}}
\acrodef{RDF}[RDF]{Resource Description Framework}
\acrodef{REST}[REST]{REpresentational State Transfer}
\acrodef{REV}[REV]{Remote Evaluation}
\acrodef{RF}[RF]{Relevance Feedback}
\acrodef{RFC}[RFC]{Request for Comments}
\acrodef{RIA}[RIA]{Reliable Information Access}
\acrodef{RMSE}[RMSE]{Root Mean Square Error}
\acrodef{RMT}[RMT]{Run Management Tool}
\acrodef{ROM}[ROM]{Read Only Memory}
\acrodef{ROMIP}[ROMIP]{Russian Information Retrieval Evaluation Seminar}
\acrodef{RoMP}[RoMP]{Rankings of Measure Pairs}
\acrodef{RoS}[RoS]{Rankings of Systems}
\acrodef{RP}[RP]{Relative Position}
\acrodef{RR}[RR]{Reciprocal Rank}
\acrodef{RSM}[RSM]{Run Storing Manager}
\acrodef{RST}[RST]{Rhetorical Structure Theory}
\acrodef{RSV}[RSV]{Retrieval Status Value}
\acrodef{RT-UML}[RT-UML]{\acs{UML} Profile for Schedulability, Performance and Time}
\acrodef{SA}[SA]{Software Architecture}
\acrodef{SAL}[SAL]{Storing Abstraction Layer}
\acrodef{SAMT}[SAMT]{Statistical Analysis Management Tool}
\acrodef{SAN}[SAN]{Sistema Archivistico Nazionale}
\acrodef{SASM}[SASM]{Statistical Analysis Storing Manager}
\acrodef{SBTO}[SBTO]{Set-Based Total Order}
\acrodef{SD}[SD]{Sequence Diagram}
\acrodef{SE}[SE]{Search Engine}
\acrodef{SEBD}[SEBD]{Convegno Nazionale su Sistemi Evoluti per Basi di Dati}
\acrodef{SEM}[SEM]{Standard Error of the Mean}
\acrodef{SERP}[SERP]{Search Engine Result Page}
\acrodef{SFT}[SFT]{Satisfaction--Frustration--Total}
\acrodef{SIL}[SIL]{Service Integration Layer}
\acrodef{SIP}[SIP]{Submission Information Package}
\acrodef{SKOS}[SKOS]{Simple Knowledge Organization System}
\acrodef{SM}[SM]{Software Model}
\acrodef{SME}[SME]{Statistics--Metrics-Experiments}
\acrodef{SMART}[SMART]{System for the Mechanical Analysis and Retrieval of Text}
\acrodef{SoA}[SoA]{Service-oriented Architectures}
\acrodef{SOA}[SOA]{Strength of Association}
\acrodef{SOAP}[SOAP]{Simple Object Access Protocol}
\acrodef{SOM}[SOM]{Self-Organizing Map}
\acrodef{SPARQL}[SPARQL]{Simple Protocol and RDF Query Language}
\acrodef{SPE}[SPE]{Software Performance Engineering}
\acrodef{SPINA}[SPINA]{Superimposed Peer Infrastructure for iNformation Access}
\acrodef{SPLIT}[SPLIT]{Stemming Program for Language Independent Tasks}
\acrodef{SPOOL}[SPOOL]{Simultaneous Peripheral Operations On Line}
\acrodef{SQL}[SQL]{Structured Query Language}
\acrodef{SR}[SR]{Sliding Ratio}
\acrodef{sRBP}[sRBP]{Session Rank Biased Precision}
\acrodef{SRU}[SRU]{Search/Retrieve via \acs{URL}}
\acrodef{SS}[SS]{Sum of Squares}
\acrodef{SSD}[s.s.d.]{statistically significantly different}
\acrodef{SSTF}[SSTF]{Shortest Seek Time First}
\acrodef{STAR}[STAR]{Steiner-Tree Approximation in Relationship graphs}
\acrodef{STON}[STON]{STemming ON}
\acrodef{SVM}[SVM]{Support Vector Machine}
\acrodef{TAC}[TAC]{Text Analysis Conference}
\acrodef{TBG}[TBG]{Time-Biased Gain}
\acrodef{TCP}[TCP]{Transmission Control Protocol}
\acrodef{TEL}[TEL]{The European Library}
\acrodef{TERRIER}[TERRIER]{TERabyte RetrIEveR}
\acrodef{TF}[TF]{Term Frequency}
\acrodef{TFR}[TFR]{True False Rate}
\acrodef{TLD}[TLD]{Top Level Domain}
\acrodef{TME}[TME]{Topics--Metrics-Experiments}
\acrodef{TN}[TN]{True Negative}
\acrodef{TO}[TO]{Transfer Object}
\acrodef{TP}[TP]{True Positve}
\acrodef{TPR}[TPR]{True Positive Rate}
\acrodef{TRAT}[TRAT]{Text Relevance Assessing Task}
\acrodef{TREC}[TREC]{Text REtrieval Conference}
\acrodef{TRECVID}[TRECVID]{TREC Video Retrieval Evaluation}
\acrodef{TTL}[TTL]{Time-To-Live}
\acrodef{UCD}[UCD]{Use Case Diagram}
\acrodef{UDC}[UDC]{Universal Decimal Classification}
\acrodef{uGAP}[uGAP]{User-oriented Graded Average Precision}
\acrodef{UI}[UI]{User Interface}
\acrodef{UML}[UML]{Unified Modeling Language}
\acrodef{UMT}[UMT]{User Management Tool}
\acrodef{UMTS}[UMTS]{Universal Mobile Telecommunication System}
\acrodef{UoM}[UoM]{Utility-oriented Measurement}
\acrodef{UPC}[UPC]{Universal Product Code}
\acrodef{URI}[URI]{Uniform Resource Identifier}
\acrodef{URL}[URL]{Uniform Resource Locator}
\acrodef{URN}[URN]{Uniform Resource Name}
\acrodef{USM}[USM]{User Storing Manager}
\acrodef{VA}[VA]{Visual Analytics}
\acrodef{VAIRE}[VAIR\"{E}]{Visual Analytics for Information Retrieval Evaluation}
\acrodef{VATE}[VATE$^2$]{Visual Analytics Tool for Experimental Evaluation}
\acrodef{VIRTUE}[VIRTUE]{Visual Information Retrieval Tool for Upfront Evaluation}
\acrodef{VD}[VD]{Virtual Document}
\acrodef{VDM}[VDM]{Visual Data Mining}
\acrodef{VIAF}[VIAF]{Virtual International Authority File}
\acrodef{VIM}[VIM]{International Vocabulary of Metrology}
\acrodef{VL}[VL]{Visual Language}
\acrodef{VoIP}[VoIP]{Voice over IP}
\acrodef{VS}[VS]{Visual Sentence}
\acrodef{W3C}[W3C]{World Wide Web Consortium}
\acrodef{WAN}[WAN]{Wide Area Network}
\acrodef{WHO}[WHO]{World Health Organization}
\acrodef{WLAN}[WLAN]{Wireless \acs{LAN}}
\acrodef{WP}[WP]{Work Package}
\acrodef{WS}[WS]{Web Services}
\acrodef{WSD}[WSD]{Word Sense Disambiguation}
\acrodef{WSDL}[WSDL]{Web Services Description Language}
\acrodef{WWW}[WWW]{World Wide Web}
\acrodef{XMI}[XMI]{\acs{XML} Metadata Interchange}
\acrodef{XML}[XML]{eXtensible Markup Language}
\acrodef{XPath}[XPath]{XML Path Language}
\acrodef{XSL}[XSL]{eXtensible Stylesheet Language}
\acrodef{XSL-FO}[XSL-FO]{\acs{XSL} Formatting Objects}
\acrodef{XSLT}[XSLT]{\acs{XSL} Transformations}
\acrodef{YAGO}[YAGO]{Yet Another Great Ontology}
\acrodef{YASS}[YASS]{Yet Another Suffix Stripper}

\end{document}